\documentclass{ieeeaccess}
\usepackage{cite}

\usepackage{booktabs}
\usepackage[official]{eurosym} 
\usepackage{float} 
\usepackage{balance} 
\usepackage{setspace}
\usepackage{amsmath,amssymb,amsfonts}
\usepackage{algorithmic}
\usepackage{graphicx}
\usepackage{hyperref}
\usepackage{textcomp}
\usepackage{color}

\def\BibTeX{{\rm B\kern-.05em{\sc i\kern-.025em b}\kern-.08em
    T\kern-.1667em\lower.7ex\hbox{E}\kern-.125emX}}

\begin{document}
\doi{10.1109/ACCESS.2018.2887201}

\title{Decision Provenance: Harnessing \\data flow for accountable systems}

\headname{\begin{footnotesize}\color{red}Published in \emph{IEEE Access}, vol. 7, pp. 6562--6574, 2019. DOI: \href{https://doi.org/10.1109/ACCESS.2018.2887201}{10.1109/ACCESS.2018.2887201}\end{footnotesize}}

\author{\uppercase{Jatinder Singh}, \uppercase{Jennifer Cobbe}, and \uppercase{Chris Norval}}
\address{Compliant \& Accountable Systems Group, Dept. of Computer Science \& Technology (Computer Laboratory),\\University of Cambridge, UK\ (e-mail: firstname.lastname@cst.cam.ac.uk)}

\markboth
{Singh \headeretal: Decision Provenance: Harnessing data flow for accountable systems}
{Singh \headeretal: Decision Provenance: Harnessing data flow for accountable systems}

\begin{abstract}

Demand is growing for more accountability regarding the technological systems that increasingly occupy our world.
However, the complexity of many of these systems --- often systems-of-systems --- poses accountability challenges.
{A key reason} {for this} is because 
the details and nature of the information flows that interconnect and drive systems, which often occur across technical and organisational boundaries, tend to be invisible or opaque. 
This paper argues that data provenance methods show much promise as a technical means for increasing the transparency of these interconnected systems. 
Specifically, given the concerns regarding ever-increasing levels of automated and algorithmic decision-making, and so-called `algorithmic systems' in general, we propose  decision provenance as a concept showing much promise.
\textit{Decision provenance} entails using provenance methods to provide information exposing decision pipelines: chains of inputs to, {the} nature of, and the flow-on effects from the decisions and actions taken (at design and run-time) throughout systems. 
This paper introduces the concept of decision provenance, and takes an interdisciplinary (tech-legal) exploration into its potential for assisting accountability in algorithmic systems.
{We argue that decision provenance can help facilitate oversight, audit, compliance, risk mitigation, and user empowerment,
and we also} indicate the implementation considerations and areas for research necessary {for realising its vision}.
More generally, we make the case that considerations of data flow{, and systems more broadly,} are important to discussions of accountability, and complement the considerable attention already given to algorithmic specifics.

\end{abstract}


\titlepgskip=-15pt

\maketitle

\section{Introduction}
Technology is increasingly the subject of public discussion and regulatory attention. 
In line with this discourse is a demand for {more} 
accountability for the technologies that now affect many aspects of contemporary life. 
This demand will likely grow as technology increasingly pervades society, particularly as visions such as of smart-cities and of the Internet of Things (IoT) come to be realised.
Transparency is important for holding those responsible for such systems to account, as it enables identification, audit, and oversight. 
Indeed, there is much current discussion in the public sphere on the transparency of major tech platforms, such as Facebook and Google~\cite{pasquale2015}.

From a technical perspective, a dominant research focus is on `algorithmic accountability'~\cite{diakopoulos2016}, where much discussion concerns issues of fairness, transparency, and explainability particularly regarding machine learning (ML) and the specifics of the decision-making elements.\footnote{For an indicative reading list, see: \url{https://www.fatml.org/resources/relevant-scholarship}.}
Generally less considered {are the challenges associated with} technology's broader operational contexts~\cite{procs}; often comprising several different technical components, perhaps managed by different entities (organisations), which come together to realise particular functionality. 
{For example, it is already common for data to flow from users via a mobile app to the app's provider, then potentially on to other third-parties (e.g. payment processors).}
In practice, this environment represents an interconnected \textit{system-of-systems}, of which data is a driver. 
The complexity of these systems-of-systems is set to increase as more and more advanced technologies are developed, deployed, interconnected, and automated, by a range of entities.
The grand visions of the Internet of Things and future smart cities are cases in point (\S\ref{interconnectedness}).

The complexity of such interconnected environments poses significant challenges for accountability. 
Importantly, the data flows that drive these interconnected systems are often invisible or opaque. This makes it difficult to exercise oversight and to determine where something went wrong and who is responsible, or in some cases, even to identify the entities involved. 
At the same time, these concerns are compounded by increasing levels of automation --- including the use of ML --- where certain occurrences can result in 
 decisions and actions with potentially immediate and far-reaching effects.

Technical measures can assist in making systems more transparent, improving the accountability of the organisations and individuals responsible for them, and assisting them in meeting their obligations.  
Towards this, there appears a clear need for provenance methods, which concern recording information about the nature and flow of data and the contexts in which it is processed; however, their use for accountability concerns in such contexts have had little consideration.

{This paper introduces \textbf{decision provenance}, which entails using provenance methods to provide information exposing \textit{decision pipelines}: {chains of inputs to, the nature of, and the flow-on effects from, the decisions and actions taken (at design and run-time) throughout systems.}
Its purpose is to assist accountability 
considerations in algorithmic systems, particularly those complex and interconnected (\S\ref{fac}).}

{In this paper, we explore how this aids accountability through assisting in regulatory oversight and technical investigation, facilitating compliance and recourse, increasing user agency, and generally contributing to better system design, operational (run-time) management, and risk mitigation.}
{We further outline the role of decision provenance in relation to ML as an illustrative case study, arguing that it will complement and support the ongoing work focusing on accountability as it regards algorithmic decision-making.}
{And we} discuss some of the implementation considerations and research opportunities 
{for turning decision provenance from concept into reality.}

Our aim is to highlight and raise awareness of the potential for information about data and its flow to help in realising more transparent, accountable, and compliant systems. 
By making the case for decision provenance, 
we seek to focus research efforts in this space.

\subsection*{Paper Overview}
{We begin by discussing the legal drivers {of} accountability and outlining the role of technology in such contexts, before setting out how the nature of interconnected systems --- an often overlooked area in accountability discussions --- poses challenges. 
We then expand on the concept of decision provenance, exploring the ways it can assist various aspects of accountability by exposing the decision pipeline. We go on to consider how this could bring benefits in a machine learning context at design-time, run-time, and in terms of post-hoc investigations. Finally, we discuss research challenges and opportunities for realising decision provenance at scale. In all, we reiterate that data flow is highly relevant to accountability discussions, and mechanisms such as decision provenance are an important complement to the considerable work that focuses on the algorithmic specifics.}

\section{Accountability}
\label{sec:acct}

Accountability involves apportioning responsibility for a particular occurrence and determining from whom any explanation for that occurrence is owed. 
Generally speaking, the entities held accountable are natural and legal persons (i.e. people and organisations), whether for their own actions or for those of people, organisations, or machines, which are under their control, or for which they are otherwise answerable.
Therefore in \textit{a systems context, facilitating accountability includes making it easier to determine which person or organisation is responsible} for a particular decision\slash action, its effects, and from (and to) whom an explanation is owed for that happening \cite{accountabilityIoT}. 

This may or may not involve exploring the inner workings of particular technologies, as tends to be the focus of the technical research community. 
While there is debate about the degree to which exposing the details of code and algorithmic models actually helps accountability~\cite{kroll}, \textit{our focus of discussion is on making transparent the connections and data flows  driving systems, so to show the context in which they are operating, their effects, and to indicate the entities involved} (of which there may be a number). 
Also relevant are technical mechanisms that assist system design, deployment and operation. These relate to accountability by facilitating compliance, oversight, and empowerment. 

It's also important to note that the person or organisation to whom a given entity is accountable will depend on the circumstances. In one context an entity may be accountable to end-users, in another it may be accountable to other systems operators, and in others to regulators, courts, or other oversight bodies. This accountability may arise as a result of statutory obligations (in data protection or privacy legislation, for example), through contractual relationships, or in 
relation to liability. When discussing accountability, it is important to bear such context in mind.

\subsection{The legal impetus}

\label{sec:legal}

From a legal point of view, accountability is often tied to notions of responsibility, liability, and transparency. Indeed, transparency is often a regulatory requirement for identifying responsibility or liability and thus facilitating accountability. Transparency can also assist in meeting other legal obligations arising from contractual relationships and elsewhere.

 `Transparency', from a legal perspective, doesn't necessarily mean full transparency over the internal workings of a system 
 (indeed, the limitations as to the benefits of full transparency have been well-identified~\cite{kroll}). 
 Instead, the transparency required in law usually involves information about the entities involved, high-level information about what is happening with data or about what systems are doing (rather than necessarily the specifics of how they function), and information about the risks of using those systems \cite{accountabilityIoT}. 

In this context, accountability may involve determining liability for an (automated) decision\slash action and, where harm arises as a result of that, what restitution is owed by who and to whom for that harm \cite{reed2016}. In a complex, interconnected system-of-systems there may also be various overlapping contractual relationships and obligations between various entities. Contractually, accountability may involve identifying where a breach of contract has occurred and thus which entity was responsible for that breach and therefore owes some kind of remedy. 

Legal requirements for accountability naturally extend beyond transparency, with compliance obligations arising through various legal frameworks. Particularly prominent are those around data protection and privacy, which is largely the focus of our discussion.
The EU's General Data Protection Regulation~\cite{gdpr} (GDPR), for example, obliges data controllers (i.e. the entities responsible for determining the means and purposes of processing personal data)\footnote{GDPR, Art 4(7).} {and data processors\footnote{GDPR, Art 4(8).}} to be able to demonstrate compliance with its various requirements, including the data processing principles\footnote{GDPR, Art 5.} if personal data is being processed. While the GDPR applies in the EU, similar data protection frameworks exist in many other countries (with the US currently a notable outlier, {though changes may be afoot})~\cite{Greenleaf}. 
{Further, the GDPR doesn't only apply to data controllers and processors located within the EU, but also to those located outside of the EU who are processing the personal data of individuals inside the EU.\footnote{GDPR, Art 3(2).}}
The GDPR therefore has a potentially global reach.

The GDPR affords several rights to data subjects (i.e. those whose personal data is being processed)\footnote{GDPR, Art 4(1).} which controllers are tasked with meeting, including the right to erasure\footnote{GDPR, Art 17.} (the `right to be forgotten') and the right to object to further processing of personal data,\footnote{GDPR, Art 21.} among others. 
It follows that transparency over data --- for example, data inventories that record where data came from, the subjects that it refers to, and where it goes to --- will be important for fulfilling data controllers' obligations in relation to these rights. It is important to note that the GDPR requires that data controllers proactively implement measures to support compliance, and particularly to meet the principles of data protection by design and by default, including to facilitate the exercising of data subject rights and to enable regulatory oversight.\footnote{GDPR, Arts 12 and 25, Recitals 59 and 78.}

Further, the GDPR places restrictions on solely automated decision-making that produces legal or similarly significant effects. This kind of automated decision-making is prohibited unless undertaken on a limited number of legal grounds. Where data controllers are processing personal data which they have obtained from elsewhere as part of this kind of decision-making, it will be important for them to know whether the applicable grounds have been met. The GDPR also emphasises the accountability aspects of automated decision-making (whether solely automated or not), particularly that which produces legal or similarly significant effects, with the so-called `right to an explanation' seeking to provide data subjects with transparency rights regarding decisions.\footnote{There is debate about the extent and utility of this `right'~\cite{Goodman2016, Wachter2017, Selbst2017}.} 

The GDPR also gives regulators the power to conduct data protection audits, and will require data controllers to establish binding corporate rules for auditing. 
And, under the GDPR, contracts between data controllers and processors must include provisions relating to the auditing of processors by controllers. 
Auditing, which necessarily requires transparency, will therefore become a key aspect of data protection regulation and compliance. 
Moreover, there are strong incentives for compliance, as the GDPR non-compliance risks serious financial penalties -- regulators are empowered to impose fines of up to the greater of \euro{20m} or 4\% of annual global turnover, {while other sanctions include corrective orders and prohibitions on processing, which can seriously impact a firm.}
Other regulations may also bear impact; an example might be the EU's proposed ePrivacy Regulation~\cite{epriv}, which as of October 2018 looks to extend certain data management obligations to non-personal data, including around electronic communications data and information relating to end-user equipment,\footnote{Proposed ePrivacy Regulation, Arts 6, 7, and 8.} which becomes particularly relevant as the IoT proliferates. 

While the GDPR is therefore a major impetus for introducing technical mechanisms to increase levels of accountability as it relates to systems, such mechanisms may also assist with the compliance and enforcement of other law and regulation. It shouldn't be forgotten that equally important in an interconnected systems context is being able to identify which organisation is responsible for a given system component (as discussed above). Having knowledge of the nature of data flow and data exchanges between systems and organisations will be crucial for accountability, as it will be of significant importance in determining liability should harms arise. Likewise, having knowledge of data flows and data exchanges would assist in fulfilling contractual obligations and in identifying who is responsible for breaches of those obligations and thus who owes what remedies to whom.

\subsection{Technical assistance}

There is a role for technical mechanisms to aid accountability.
It is said that \textit{at a technical level, accountability is grounded in transparency and control} \cite{weitzner2008, accountabilityIoT}. 
Technology will not itself `solve' accountability issues, and as mentioned earlier, technical transparency might not itself address legal accountability concerns; however, \textbf{technology can provide the tooling and information to assist and facilitate accountability}, for example, by providing `evidence' indicating what is happening, or by enabling constraints, responses, and rectifications.

Such  technical transparency could, for instance, provide knowledge of the legal entities involved in a system and their roles, {which otherwise may have been invisible from various perspectives.} This has accountability implications as explored above.
Further, and as well as allowing those responsible for systems to be held to the standards required by law, transparency providing knowledge of the nature of systems 
can help in evaluating those aspects for correctness, bias, fairness, and whether they are operating within the parameters that are expected or desired by their designers, operators, or end-users (\S\ref{sec:ML}). 
Transparency thus also facilitates control through intervention, where appropriate, so that where problems or potential issues have been identified within or resulting from a system, the designer or operator of the system can intervene (automatically or manually) to address them, while also enabling users and overseers to set preferences or constraints (technical or otherwise) to ensure that the functionality accords with their requirements. 

In other words, though technical measures can assist the more formal aspects of holding one to account, they also relate to broader notions of accountability by better allowing those responsible to meet their obligations, and by facilitating oversight, intervention and agency.
However, a highly interconnected context poses particular accountability challenges which to date have been underexplored. We consider these next. 

\section{The challenge of interconnectedness} \label{interconnectedness}
\label{sec:intercon}

Accountability is often discussed in relation to a single organisation, or in a systems-context, as discrete systems, where the entities involved are known or predefined; e.g. concerning the actions undertaken by a particular organisation.
However, the increasingly interconnected nature of systems means that in many cases they do not operate discretely and in isolation but are employed as part of a \textit{system-of-systems}~\cite{maier1998,sysofsys}, with potentially many entities involved~\cite{gubbi,twentycon}.

One system, for example, may take inputs from a range of sensors and may then produce an automated decision in the form of a data output that itself forms the input to another system, which, in turn, may produce an output through a device that creates a physical interaction with a real-world object or a person; emerging autonomous transport systems will likely entail lots of such  functionality~\cite{gerla}. 
In other words, the individual component systems interact and combine to bring functionality through a system-of-systems, where components may be potentially managed by multiple different entities. 

\begin{figure}[thp]
\includegraphics[width=\columnwidth]{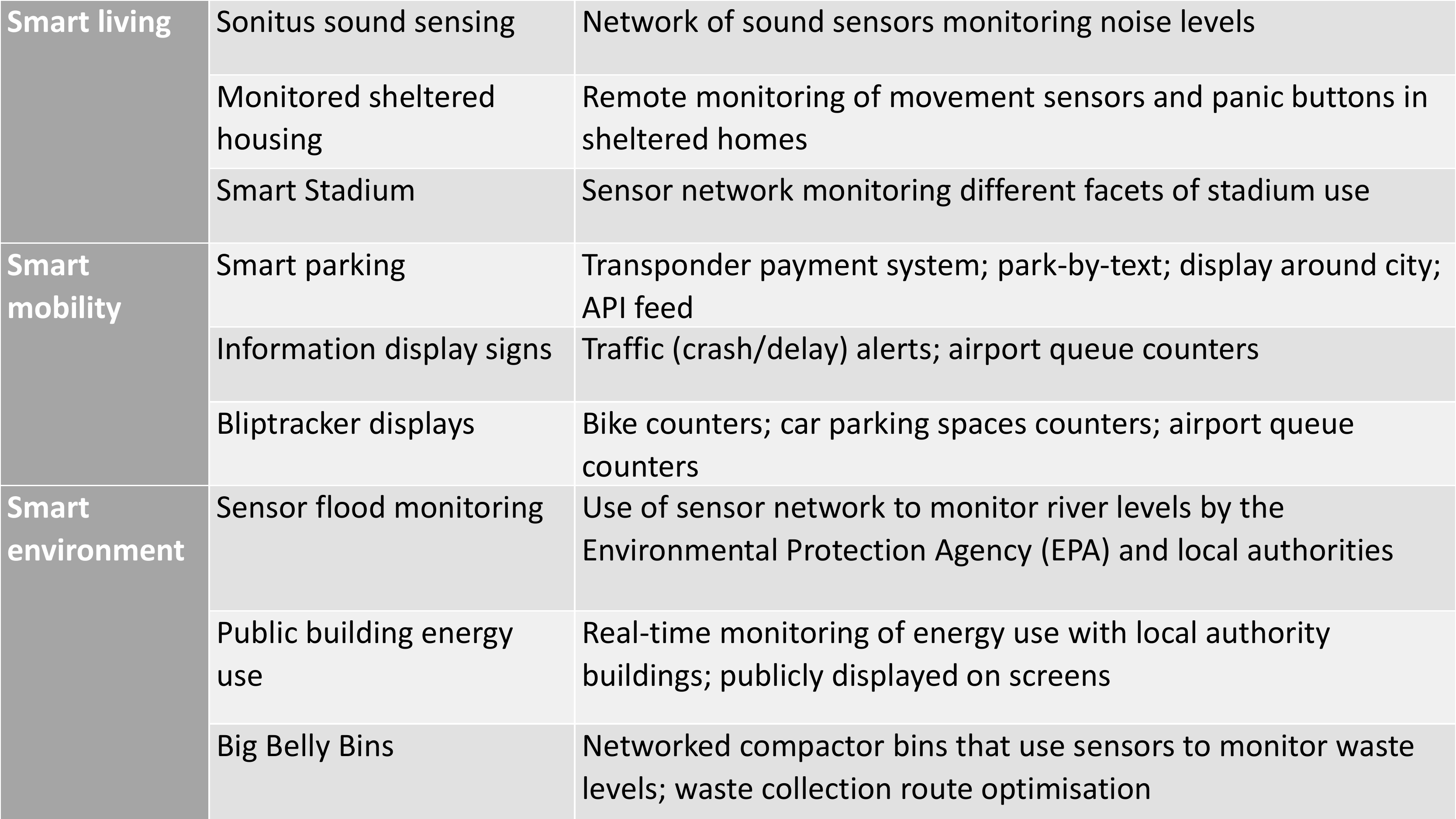}
\caption{ Some smart city technologies implemented in Dublin as of  2016~\cite{kitchin2016}. The broader visions for smart cities is for systems like these, as well as new systems, to interact with each other as well as with various commercial and consumer-oriented services.
\label{fig:sc}}
\end{figure}

The general direction towards more connectivity can be seen, for example, in relation to smart cities~\cite{dijk2015, kitchin2016, mehmood2017}. 
As Fig.~\ref{fig:sc} shows, there are already a number of smart systems which are being deployed to manage cities. 
And, as the vision of the smart city is increasingly realised, increasing numbers of complex systems will be deployed and will interact with each other. 
At a more granular level, we already see systems which are composed of interconnected systems -- the popularity of cloud \textit{([something]-as-a-service)} a case in point.

\newlength{\xfigwd}
\setlength{\xfigwd}{\columnwidth}

Importantly, in these interconnected systems \textbf{data flow drives everything}, in that the flow of data between, through, and across systems works to integrate them and brings their functionality~\cite{mwbi}.
A given data flow  might encapsulate, for example,  a sensor reading, the contents of a web form, \textit{the inputs to or outputs (decision or actions) from a machine learned model}, a batch transfer of a dataset,  an actuation command, a database query or result, and more. 
However, in most cases \textbf{the visibility over the nature and flow of data is lost once it moves beyond a particular boundary}, whether that is technical (e.g. between software components and services) or administrative (e.g. between organisations or jurisdictions).

Likewise, the lack of visibility over data flows in interconnected systems poses various challenges from a legal point of view. 
Not only will it be important to identify where data has been obtained from, but in order to meet data protection obligations it will be important for data controllers to know under what circumstances and on what conditions personal data was obtained. 
Similarly, if data controllers share data with others, it will be important for each entity to have a record of when the data was shared, from and to whom the data was shared, the basis on which the data was shared, and any conditions attached to the use of that data. This is all in addition to information regarding how the data is managed and processed.

For example, the GDPR requires that personal data be collected for specified, explicit, and legitimate purposes and only be processed in a manner compatible with those purposes (a principle known as `purpose limitation').\footnote{GDPR Art 5(1)(b).} 
Knowledge of where data has been obtained from and under what circumstances it was obtained {will also likely} be important {for complying with restrictions in other regulations, such as the} 
proposed ePrivacy legislation on the use of various kinds of non-personal data, as well as for establishing liability where harms are caused.\footnote{{Though, as mentioned, ePrivacy Regulation is still but a proposal, it is indicative of the sorts of concerns we can expect to see in going forward.}}

However, meeting these kinds of obligations may be difficult, and in some cases  impossible, without some knowledge of the provenance of incoming and the path of outgoing data flows. 
To return to the data protection example, where data controllers have obtained personal data from a third party but lack knowledge of the conditions on which that data was obtained by that third party, such as that the data is not to be used for the purposes of advertising, complying with the kind of restrictions imposed by the principle of purpose limitation  would be particularly difficult. 

{The general opacity of the interconnections between systems,}
coupled with the general lack of technical means for tracking data flow at such a scale, therefore poses accountability issues, as it becomes difficult to discern the technical components involved, and then who is responsible for these~\cite{mwbi,pasquier2017}. 
Moreover, decisions and actions somewhere within an assemblage might propagate widely, making it difficult to trace the source (organisational or technical) when concerns arise, and all the consequential (flow-on) effects. 
As an example, a reading from a faulty sensor could lead to cascading knock-on effects causing significant disruption, though its relationship to this may not be readily evident given the gaps in `time and space'. 
 
\section{Decision provenance: exposing the decision pipeline} \label{decprov}

\label{sec:pipeline}

\begin{figure*}[!t]
\centering
\includegraphics[width=\linewidth]{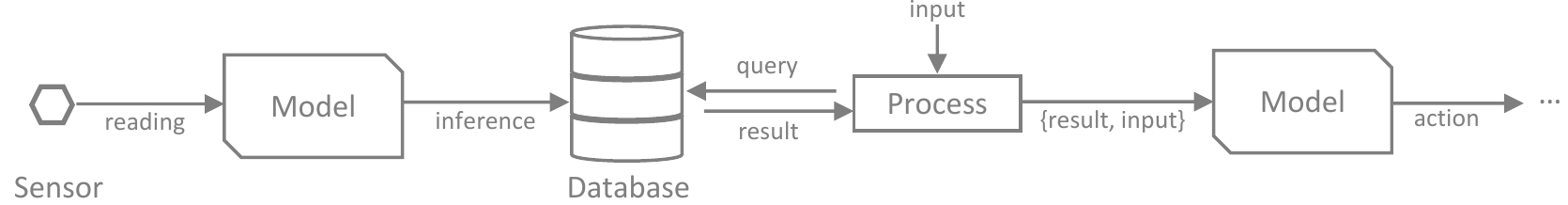}
\caption{ 
{An example illustrating an interconnected system-of-systems. Here, sensor}
data drives the inferences of a predictive model which are recorded in a datastore. 
This is then queried by a process which feeds the result, alongside another input, into another model that affects an action.
One can imagine similar arrangements involving, for instance, a smart home management system, which  automatically controls the environment, being influenced by systems that forecast weather, wearables inferring a resident's mood and well-being, and so forth, where each component might be managed by a different entity.}
\label{fig:systems}
\end{figure*}

From the above, we can see that greater visibility over the {interrconnections and assemblages} of systems is important for increasing accountability for them. 
Given that data drives systems, there appears to be real potential in applying data provenance methods as a means of working towards this.

{ 
Put simply, data \textit{provenance} concerns capturing information describing data: recording the data's lineage, including where it came from, where it moves to, and the associated dependencies, contexts (environmental and computational), and processing steps~\cite{provprimer}.\footnote{A detailed overview and elaboration of the nuances of data provenance can be seen in~\cite{provprimer,melProv}.} 
It is commonly used within research to assist reproducibility by providing records of data, workflows and computation~\cite{melProv,davidson2008}.}

Provenance is an active area of research~\cite{melProv}, and is commonly applied in a research context to assist in reproducibility by recording the data, workflows and computation of scientific processes~\cite{melProv,davidson2008}. 
The potential for provenance to assist compliance with specific information management obligations has previously been considered (often focusing on a particular technical aspect, be it representation or capture)~\cite{aldeco2010,ieeecloud,tan2013,mwbi,pasquier2017}. 
However, the use of data provenance methods generally for systems-related accountability concerns, as detailed above, has yet to be considered in depth.
As such, here we first motivate the potential for provenance to assist algorithmic accountability, and later highlight research opportunities for moving forward.

\subsection{Decision Provenance}

Given (i) the concerns around accountability for automated and algorithmic decision-making systems, 
(ii) the increasingly interconnected nature of system deployments, and
(iii) the potential of provenance methods in this space,  
we propose \textit{decision provenance} as a way forward.\footnote{Distinct from the DECProv ontology for the general modelling of the specifics of a decision-making process: \url{https://promsns.org/def/decprov}.} 

Decision provenance 
concerns the use and means for provenance mechanisms to assist accountability considerations in algorithmic systems. 
 Specifically, \textit{\textbf{decision provenance}}
involves providing the information on the nature and contexts of the data flows and interconnections leading up to a decision or action, the flow-on effects, and also how such information can be leveraged for better system design, inspection, validation and operational (run-time) behaviour.
In this way, decision provenance helps expose the \textit{decision pipelines} in order to make visible the nature of the inputs to and cascading consequences of any decision or action (at design\slash or run-time), alongside the entities involved, systems-wide (e.g. see Fig.~\ref{fig:systems}).
The broad aim 
 is to help increase levels of accountability, both by providing the information and evidence for investigation, questioning, and recourse, and by providing information which can be used to proactively take steps towards reducing and mitigating risks and concerns, and facilitating legal compliance and user empowerment. 

\subsection{Provenance: facilitating accountability}
\label{fac}

As we have outlined, the opacity of 
 the interconnections in complex systems poses significant accountability challenges. 
Decision provenance concerns capturing records of data flow throughout a system, as relevant for accountability considerations; which can include the nature of who (legal entity) the data comes from or goes to, how the data was processed, used, and other appropriate contextual information, such as data protection aspects, system configuration, actions of staff, etc. 
This works to expose the decision pipeline by providing records regarding: \\ \indent(i) the history for particular data, and \\ \indent(ii) the broader view of system behaviour and interactions and the interactions between entities. 

\noindent Such records are not only useful for (ex-post) audit and investigation, but can also actively drive or determine interventions and actions.\footnote{In a similar vein to how run-time provenance has been used for detecting and notifying of  system faults~\cite{frap}.}

As a result, decision provenance can help improve accountability by assisting oversight, empowerment, investigation, audit, and compliance. There are potential benefits for system designers, operators, auditors, regulators, and others from both a technical and a legal point of view, as well as for end-users. Generally speaking, more information regarding the nature of systems could allow those building and deploying systems to improve their quality and to identify points of failure, while empowering others by offering more insight into the nature of the systems and data that is beyond their direct control~\cite{mwbi,accountabilityIoT,pasquier2017}. 
Proactive accountability measures can be employed, where the knowledge regarding nature of the data flowing into, within, and beyond a system drives or determines particular actions or behaviour, e.g. influencing if and how data is processed. 
Responses may be human or organisational, perhaps automated through a triggered or event-based response~\cite{mwbi}.

The general idea is conceptually similar to the tracking of physical items through product supply chains, which helps manufacturers gain a better understanding of the provenance of their products and of the materials therein \cite{new2010}. 
This allows factories and warehouses to respond and be accountable to manufacturers, which in turn allows manufacturers to be accountable to regulators and consumers. 
{Naturally, provenance will not solve all accountability challenges, but it assists by providing visibility in complex environments by exposing the relations and dependencies where such visibility may not otherwise exist.}

\subsubsection*{Accountability benefits}

We now explore some ways in which decision provenance can assist specific accountability considerations of algorithmic systems, before the next section that explores ML-driven systems 
{as a case study}.
It is worth noting that while we have delineated these benefits, they can be interrelated -- legal investigation, for example, may involve technical audit and investigation.

\subsubsection{Understanding system  behaviour}

{Generally, knowledge of the nature of the data flowing in, out and within a system, in terms of where it has come from, where it goes, how it has and is being used, as well as the relevant environmental and computational contexts, }can help systems designers and operators (manually or through automatic means) monitor, maintain and improve the quality of their systems on an ongoing basis. 
The potential for provenance methods to allow one to `see' across systems and their interconnections can also enable proactive steps to be taken in order to address problems before they arise. 
This works towards increasing the overall quality, security, and reliability of complex, interconnected systems and services, {which has clear organisational benefits. } 

\subsubsection{Compliance and obligation management} \label{compliance_ob_management} 

Decision provenance can assist those responsible for systems in complying with their legal (and other) obligations~\cite{ieeecloud}.

As discussed in \S\ref{sec:acct}, servicing many of the rights afforded to individuals in data protection law could be facilitated through mechanisms for tracing the paths of data so as to identify where data resides, with whom it was shared, and how it was processed (e.g. see \cite{ieeecloud,pasquier2017}). 
Decision provenance could work to create data inventories, which can be of use in meeting obligations such as the right to erasure, subject access requests, and others in data protection law. It would make it easier to meet obligations to provide information to data subjects, including on what personal data is being processed, where it has been obtained from, and which other data controllers or processors will or have received the data. 

Decision provenance could also allow for the tracking of conditions associated with the processing of personal data, such as those relating to consent for processing data, the purposes for which data may be processed, and the sharing of data with other entities.
This would assist data controllers in complying with the various data protection principles. This may also be of use in assisting compliance with future regulations, e.g. the proposed ePrivacy Regulation, which establishes some similar requirements on the use of some non-personal data.

Decision provenance can similarly assist in managing a broader range of legal, regulatory, and other obligations (`soft law', best practices, etc).  
For instance, where contractual obligations exist between entities, knowledge of the nature of data flow between parties could make it possible to ensure and validate that data is processed in a way compatible with the contract governing that data flow or processing relationship. 
And information of the sources and lineage of data used for analytics and ML (see \S\ref{sec:ML}) might assist with issues of unfairness and discrimination~\cite{wef}.

Reactive, event-based mechanisms  could also automatically take actions to assist compliance and obligation management~\cite{mwbi}.
For instance, provenance information could be used to trigger particular compliance operations; such as to automatically report data breaches to the relevant authorities; to screen and filter out data based on compliance criteria (e.g. data past an expiry date or use); to not act on inputs that may have come from, or through, an unreliable entity~\cite{ieeecloud}; or to automatically prevent data flows that are unexpected (in terms of a pre-defined management policy~\cite{mwbi}).

\subsubsection{Oversight and regulatory audit}

The use of decision provenance techniques has much potential to aid the auditing and oversight activities of regulators, as provenance involves generating detailed information for assisting oversight that was previously unavailable. 
This might include, for instance, data protection regulators using provenance data to assess that the data flows pertaining to an organisation (data controller) are appropriate, or more generally that the organisation has the appropriate data management (and other) practices in place, and has taken the appropriate actions regarding specific incidents. 

Beyond regulators and oversight bodies, the same methods are useful for organisations to help in managing their own data `supply chains'.
For example, such information assists an organisation in evaluating that other business they engage with (e.g. through a data processor relationship) are meeting their contractual and data management obligations by handling data appropriately.

\subsubsection{Technical audit and investigation}

Decision provenance can also help facilitate technical audit and investigations in complex systems by helping systems designers and operators to identify 
points of failure or {issues} requiring further investigation. This includes both post-hoc investigation as well as run-time reaction (e.g. alerting of unexpected  system behaviour). 
Exposing the decision pipeline can help in identifying aspects such as the steps behind the leakage of personal or otherwise sensitive data, and in tracking the cascading consequences of a particular (perhaps erroneous) event.
This could, for example, 
make it easier to determine whether it was a particular data source such as a sensor producing invalid readings, poor data selection by a system designer, or a learned model being inappropriately applied, that eventually led to decisions with poor outcomes being made. 

The importance of such concerns will only increase as the visions of the IoT and smart cities become a reality, due to their complex and highly-interconnected nature and also their potential for emergent functionality and real-time reconfiguration~\cite{accountabilityIoT,mwbi}.

\subsubsection{Liability and legal investigation}

Where harm is caused by a failure, particularly in complex systems environments,
decision provenance can help identify which system caused that harm, thereby helping to identify the organisation responsible and potentially liable for that harm and assisting in holding them to account. Provenance data may also work to absolve systems designers and operators from responsibility by providing evidence demonstrating that the right decisions were made and actions were taken. In this way, decision provenance is useful for organisations\slash operators, to help manage their liability concerns, as well as for regulators and law enforcement.

More generally, decision provenance can assist in ascertaining whether legal obligations were met. For instance, where contractual obligations exist between entities, if there is a breach of contract, knowledge of data flow assists systems operators in identifying where that breach occurred and therefore who is responsible and what remedy is owed. 

In this way, decision provenance can offer preemptive advantages, by allowing organisations to have greater control over how their data (that for which they have responsibilities) and services are subsequently used.
For example, organisations could contractually require visibility over how certain services and data are used by external parties, where decision provenance records provide such information.
This could limit instances of data misuse by giving organisations a technical means to monitor how their data is used, possibly in real-time, and enable action where such uses are deemed unacceptable.
Indeed, enhanced means for governance could encourage collaboration, and help in tackling data silos.

Similar methods could be used to investigate suspected or actual breaches of other legal\slash regulatory obligations. 

\subsubsection{End-user agency and empowerment}

There are also potential benefits for individuals in terms of increased agency. 
Decision provenance information could help users make better-informed decisions about the services that they use. 
If  a user could see in advance that giving data to a particular system could, for example, (a) result in certain decisions being made that have particular undesired consequences; or (b) flow to an undesired entity (such as a certain advertisement network), then this helps that user to make a more informed choice about whether to use (allow their data to flow into) that system. 

Of course, evidence of prior behaviour does not necessarily imply future behaviour will remain the same; however it can be indicative. That said, proactive measures could also be employed to allow users to set policy that constrain particular information flows~\cite{mwbi}, or keep users informed of any change in circumstance. For instance, mechanisms could be built whereby the detection of unexpected flows could trigger an alert indicating a change in policy or system operation (or a potential data leakage), e.g. a social media firm suddenly engaging a new advertising network. 
Such measures, driven by provenance information, could allow users to take action, e.g. exercise their rights, complain to a regulator, stop using the service.

These aspects are important given the ever-increasing prominence of data-gathering systems, and in addition to generally increasing levels of agency, might facilitate a more informed and effective exercise of data subject (and other) rights. 
Measures that help users make better-informed decisions about which systems they engage with may also work to incentivise those producing and operating technology to improve their practices at all steps, and to help better align their offerings with user expectations and preferences.

\section{Case study: machine learning pipelines}
\label{sec:ML}

A key driver for discussions of algorithmic accountability, automated decision-making, and indeed, decision provenance, is the increasing prevalence of machine learning (ML). 
The popularity of ML, which drives algorithmic decision-making, has been a core instigator of discussions regarding `FAT' (fairness, accountability, transparency).\footnote{See, for example, \url{http://www.fatml.org}.} 
We see decision provenance as a key component within the wider context of fair, accountable, and transparent technologies, and discuss ML as an example use-case to illustrate how it could help advance such aims.

ML is driven by data. It works to uncover patterns in data so as to build and refine representative mathematical models of that data which can be used to make predictions, decisions, and gain knowledge and insight~\cite{james2013}.
These models can be used for problems that would otherwise be challenging to program specific rules for, such as object classification in images, by instead having the model derive these through trends in the training data. 
In a systems context, these models are applied to new (`live') input data, with the corresponding output(s) (also data) representing a prediction, a decision, an action, an inference, and so on. 

ML raises interesting considerations in a systems context. 
First, ML is part of the `big data' trend, where vast quantities of data which can originate from a wide variety of sources is relevant for both the building and use of ML models. 
Yet, the data itself can often be the problem; biased data will often result in biased models~\cite{hajian2016, kamiran2012, kirkpatrick2016, pedreshi2008}.
This is particularly problematic when consequences and implications are significant, such as with criminal justice risk scores and credit offers~\cite{kirkpatrick2016}, and the increasing popularity of ML risks those who are not domain experts unintentionally misusing datasets which are unsuitable in particular contexts~\cite{gebru}.

Further, we increasingly see ML models being offered as a service,\footnote{See, for example, services currently offered by Google: \url{https://cloud.google.com/products/machine-learning}, Microsoft: \url{https://azure.microsoft.com/en-gb/services/cognitive-services}, and Amazon: \url{https://aws.amazon.com/machine-learning}.} allowing users to pass input data to these services and receive their predicted outputs. 
In practice this means that ML models can be integrated together with other software processes, and indeed other models to form complex, interconnected chains of systems (see Fig.~\ref{fig:systems}).

Given the ever-increasing complexity of algorithmic systems, and as systems ascend beyond the scope of single organisations and legal jurisdictions, \textit{information of the source, lineage and nature of data, models, and decisions\slash actions taken} {will also be important even when considering accountability and transparency in relation to the decision-making elements (ML, analytics).}
As is the case with interconnected systems in general, data flow is the enabler of ML. 
This includes, for example, the data flows relating to the training of models (design-time), the use of models that make decisions (run-time), and the flow-on effects of those decisions throughout the system.
As such, we argue that mechanisms for decision provenance, that operate throughout, is becoming ever more important for assisting accountability in these systems arrangements. 

\subsection{Assisting training and design}  

\label{dataset_construction}
Machine learning models are built (trained) on data, and therefore reflect the nature of that data. 
Training models on data of a poor quality (e.g. unrepresentative, biased,  erroneous, etc.) can lead to issues being encoded within these models and reflected in the model's application. 
Indeed, models can encode issues of bias, discrimination, and unfairness {that are inherent} 
in that data~\cite{hajian2016, kamiran2012, kirkpatrick2016, pedreshi2008}.
In an interconnected context, a training dataset may comprise  data from a range of sources, including sensors, human input, system logs, data brokers, and so on, and can also include outputs from calculations, analytics, or indeed, other ML models (\S\ref{interconnectedness}). 

\begin{figure}[H]
\centering
\includegraphics[width=0.9\linewidth]{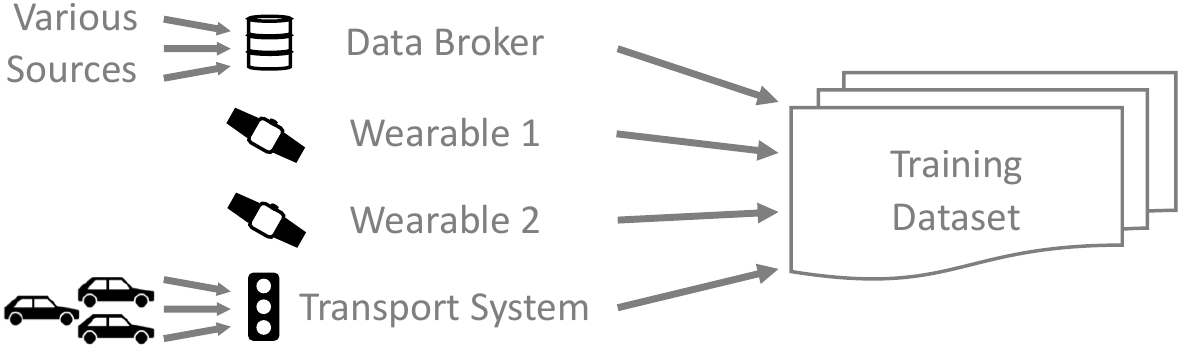}
\caption{Training data comprised from multiple sources.
}
\label{fig:datasets}
\end{figure}

Provenance has a clear role to play in showing the nature of the training datasets, in that revealing the pipeline can allow for greater knowledge of the sources of data, assisting in assessing their veracity.
Such oversight could highlight the potential for errors, inherent bias, or unfairness present in any models which use this data, enabling ML practitioners to be better informed and take redeeming actions.
An example of this in practice is the careful use of sampling techniques and respondent weighting in opinion polling.\footnote{And indeed, such provenance records also facilitate the use of post-hoc analysis on such polls (e.g.~\cite{mellon}), akin to what we discuss in \S \ref{investigation}.}
This use is also in line with common applications of provenance for research reproducibility.

It is increasingly recognised the importance for the ML practitioners, themselves, to record information relating to the datasets, such as their motivation for creation, composition, process of collection, preprocessing, distribution, maintenance, and legal and ethical considerations~\cite{gebru}.
Such records can be used to evaluate whether certain data should be permitted for use in the construction of a ML model, based on the metadata of that dataset.
While current steps entail manual (human) processes to record such information, typically within the scope of a single organisation~\cite{gebru}, decision provenance involves technical means for collecting information relating to data and models, potentially including the computational contexts in which models were constructed, right 
throughout a data supply chain. 
In this way, decision provenance can provide much information, perhaps complementing other records, for data and model assessment.

\subsection{Assisting operation}

Machine learning models are often applied to `live' data, which is given as an input to a trained model in order to produce {one or more outputs. 
These outputs may result in physical-world actions (i.e. an actuation command) or outputs that feed into other datasets or trigger other models, possibly crossing administrative boundaries and affecting a range of entities (see Fig.~\ref{fig:systems}).
By capturing the provenance and nature of the model inputs and the decision itself (i.e. the model outputs), 
decision provenance 
could be used to detect conflicts and uphold compliance at run-time.
}

As an example, {provenance information could inform model selection: where 
 the lineage of data determines (at run-time) the appropriate model on which the data is applied. 
 For example, if data being passed into a live model comes from a recognisably different source (sampling distribution) than the data that was used to train the model,  actions could be taken to prevent such processing. }
Further, rules could, for example, prevent a data subject's personal data being presented to a ML model --- at run-time --- unless they have chosen to allow solely automated decision making\footnote{GDPR, Art 22.} (which itself could potentially also form part of the provenance, e.g. recorded against the data, or evident based on the chain of components surrounding that data).
The above represents but a few examples of how decision provenance would facilitate a number of applications and advantages that {assist with algorithmic accountability\slash decision-making}.

\subsection{Using the decision pipeline: points for investigation} \label{investigation}

As \S\ref{decprov} describes, decision provenance helps investigate complex and interconnected systems by exposing the data flow and interconnections leading up to a decision or action, and the entities involved. 
Once the decision pipeline has been used to identify the particular source(s) (technical components and\slash or legal entities) \textit{potentially} related to an issue, auditors can then narrow their focus and go into detail. For instance, {if the issue concerns a decision originating from a specific ML model} in the chain of systems,  this may involve going on to inspect the workflows and processes around the construction of that model~\cite{procs}, e.g. which might entail looking at the `dataset's datasheets'~\cite{gebru}, or by using techniques discussed in the algorithmic accountability literature (e.g.~\cite{diakopoulos2016}).
Information about the workflow of model construction and deployment itself facilitates auditing of, for example, the learning algorithm (or algorithms in ensemble models) used, hyperparameters, predictor variables, and coefficients (variable weights)~\cite{schelter2017}.

As such, decision provenance complements the ML-oriented work on `algorithmic accountability' { (i.e. that focused on the decision-making elements),} by providing auditors with information of the wider, practical context in which automated decision making processes were defined, built and operate, as a means for indicating where deeper (model-centric or other) investigations can then take place. 

\section{Moving forward: Implementation considerations and research opportunities}
\label{futurework}

{
Despite the potential for decision provenance,  work is needed to make it a reality.\footnote{That said, there are real and immediate benefits in employing {methods for capturing provenance information \textit{now}, in more local scenarios and as much as is possible with current tooling (technical or otherwise),} as even internal organisational usage can help in managing systems, processes, and obligations, while assisting compliance and providing evidence demonstrating good practice~\cite{ieeecloud}.}
Though, as mentioned, provenance has been considered for some specific compliance aspects (see for e.g.~\cite{aldeco2010,ieeecloud,tan2013,mwbi,pasquier2017}), the broader accountability considerations have yet to be explored in depth. 
In terms of decision provenance operating between systems and across boundaries, many of the open challenges predominately stem from the scale, federation, and complexity of what are effectively wide-scale distributed systems that can encompass a range of technologies, and also a number of organisations with different and possibly competing incentives. For provenance to enable accountability, crucial is the consistency, visibility, and veracity of provenance information across this federated environment.}

Towards this, we now indicate some areas for consideration, focusing on the methods for gathering and representing provenance information, as that provides the foundation for the accountability opportunities previously discussed (\S\ref{sec:pipeline}). 

\subsection{Means for appropriate capture and record management} 
\label{sec:capture}

A key consideration are the mechanisms for capture, i.e. the technical means for producing provenance information. 

One categorisation of provenance mechanisms entails two categories~\cite{provprimer}: \textit{disclosed} and \textit{observed} provenance. 
Disclosed provenance tends to be application-oriented, where the details of what to record, and when, are written into application(s) code (often in APIs, as points of data exchange). 
This allows customisation by the designers as to what is deemed important, but will only capture what is explicitly programmed for{, and depending on the implementation, may facilitate users (or others) providing extra `manual' provenance information relevant to provenance records.}

Observed provenance captures information by observing what occurs; for example, by embedding the capture mechanism into the platform\slash infrastructure (e.g. an operating system (OS)), to capture the data flows regarding the applications running within that environment~\cite{socc,camflow,spade}. 
By the capture mechanism operating externally from applications, this allows the interactions between various applications (system components) to be recorded irrespective of the application specification and design -- without {the need for} developer (let alone user) intervention or even knowledge. 
Further, potentially capturing every data flow in the operating environment reduces the propensity for `missing something'. 
However, a general observed capture mechanism is comparatively less-targeted, and raises serious overhead considerations given the propensity for the volume of the recorded information to be extremely large and complex~\cite{ifa}.

{
In practice, it is likely  that decision provenance may require a combination of capture mechanisms to ensure that  the appropriate information is recorded~\cite{provstack,layering},\footnote{Though this could be limited where provenance concerns are narrow and focused.} as, for instance, some aspects might be observable (e.g. data flowing between software components in a cloud service), and others rather unobservable from purely a technical perspective and therefore requiring more manual intervention (e.g. capturing the intentions or thoughts of employee users).}

An interconnected system-of-systems --- with its many moving parts --- poses practical implementation challenges, given the requirement for the provenance regime to operate consistently throughout, as we now explore.

\subsubsection*{Capture mechanisms}
Capture mechanisms should be able to {function} 
throughout the entire decision pipeline, by operating at the appropriate points within and across technical and organisational boundaries. 
Capture regimes that are bound to a specific application might suffice where the pipelines\slash supply chains are pre-known, well-defined and fairly static; though of course, such approaches are limited to only those pre-defined applications and organisations.
In contrast, observed approaches suit capturing detailed information across applications, but only within the scope of specific operating environments (e.g. a platform\slash OS instance).
There is on-going work towards  the challenges in maintaining consistent capture regime across boundaries, administrative\slash organisational and operational, e.g. as data flows from one OS (machine) to another~\cite{pasquier2017,spade,mwbi,socc,tan2013}.

In practice, it is likely that exposing the decision pipeline will require several complementary capture regimes. 
This accords with the conceptual provenance stack, which defines a series of layers for focusing specific provenance considerations --- corresponding to different levels of abstraction --- to enable a more complete regime~\cite{provstack,layering,melProv}. 
{Work will be needed to ensure interoperability and\slash or to enable the reconciliation of records across different capture mechanisms.

In line with this, an important area of work is in defining the standards, guarantees and best practices regarding capture mechanisms. 
{
These will be important for increasing levels of completeness in provenance records, including the reconciliation and interoperability just mentioned, and  more generally, to enable provenance at scale. They will also assist in (i) enabling ex post accountability, i.e. enabling investigation, as well as providing (ii) a foundation for enabling proactive (ex ante) responses (\S\ref{fac}), including technical policy enforcement regimes that can react to certain events based on the provenance information~\cite{mwbi}.}}

\subsubsection*{Data management}
Related is how the captured provenance data is recorded and made accessible across boundaries (technical or organisational). 
The storage requirements and possible sensitivity of provenance data tends towards such information being federated across organisations. 
If provenance information is federated, questions arise as to how it can be reconciled across {systems and organisational boundaries (i.e. in addition to across the technical boundaries mentioned above)} in order to make the decision pipeline visible, and how this provenance data is to be aggregated (if at all), queried, and so on. 

Another consideration are the record-keeping requirements: for how long should provenance data be kept?
Data retention considerations are particularly important if serving as `evidence' supporting investigations or other accountability processes. 
There is scope for methods that aggregate, summarise and compress such information to assist storage\slash management overheads.

Further, the security of, and access to the provenance data itself raises accountability questions, given that, as information about data, systems and processing, it may itself be sensitive~\cite{pasquier2017}, {or come with various legal obligations -- particularly in a cross-organisational context.}\footnote{{Note that the nature of provenance information may well serve to place additional legal, management and risk-related obligations on organisations.}} 
Secure management of this data is one area of consideration~\cite{secprov}, including determining the contexts in which provenance information can be shared, and access controls might be restricted or opened to others, such as regulators, in certain circumstances.
In such scenarios, distributed queries that operate across provenance repositories may offer promise. 

The above concerns are context-specific, depending on the nature of the record and what it holds; e.g. some records may be more useful or sensitive than others. 
In practice, it may be that {data management and reconciliation decisions are} driven by {standards, best practices, on the guidance of overseers, or determined by legal, regulatory or contractual obligations}{, presenting potential ways forward}.

\subsubsection*{Trust}
For provenance information to be useful, it must be reliable. 
However, this is a particular challenge here for several inter-related reasons. 
First, the risks and incentives in an accountability context are complex, given that the provenance information relates to responsibilities  from which onerous consequences might result.\footnote{There are, however, incentives to implement such provenance mechanisms, for instance, to produce evidence that aid arguments for  absolving responsibility (\textit{``I've done nothing wrong''}).} 
Second is the inherent federation in terms of the mechanisms for capture and what is recorded. 
Third, the nature of the means for capture that are used, e.g. disclosed or observed and the level of the technical abstraction in which they operate, impacts issues of reliability, validity, accuracy, usefulness and completeness. 
These aspects in combination raise issues of trust in multi-party scenarios, where data flows across administrative or organisational (i.e. responsibility) domains. 

In all, there is a clear requirement for means that assure and ensure the integrity of the capture regime, recorded data, the query mechanism, and any other compliance mechanisms (such as policy enforcement regimes) that leverage provenance data, and also  to ensure  that this integrity is maintained throughout any pipeline~\cite{moyertrust,pasquier2017,mwbi}.
Towards this, standardisation, verification, attestation, and secure logging mechanisms will all be relevant. 

\subsection{Making provenance data meaningful}

{Regardless of the capture and storage mechanisms, provenance records should be consistent and meaningful in order to be useful in assisting accountability concerns.}

\subsubsection*{Representation of provenance data}
In a systems context, it is important that there is a common representation of provenance data so that it can be widely understood and interoperate system-wide. 
Standards will help support the interpretability and understanding of audit records within and across systems.
The W3C PROV\footnote{\url{https://www.w3.org/TR/prov-overview/}.} standard provides an extensible\footnote{For instance, see the PROVOne extensions for scientific workflows: \url{https://purl.dataone.org/provone-v1-dev}.} mechanism for modelling provenance data, while ontologies (description vocabularies) provide the means for describing what has been captured. 
W3C PROV is capable of capturing aspects relevant for compliance and accountability, for example, for recording which parties undertook various activities with regard to data, or modelling GDPR compliance~\cite{ujcichprv}. 

In moving forward, one consideration is whether specialised models or vocabularies will be necessary for accountability purposes (be they general or specific), as well as to help align the provenance information that is captured at different technical layers of abstraction (\S\ref{sec:capture}). 
Different extensions may also be needed to cater for the specifics of a particular application domain; what needs to be captured for an automated traffic management system will differ to that of an e-commerce website.

\subsubsection*{Usability of captured data}

A wider consideration is how stakeholders from diverse backgrounds and with varying goals can interpret and use provenance data; though this is a general challenge for provenance given that captured data is often, or can quickly become, extremely complex~\cite{chen2012, davidson2008, oliveira2017, wang2018}.
Such concerns are exacerbated in a large-scale systems context, and are problematic where accountability is the aim. 
Some (e.g. end-users) may be interested simply in the entities involved, while others  (e.g. regulators) may require more information. 
This means that the provenance data captured may be difficult to interpret for some --- for instance, making sense of data representing different levels of the technical stack\slash abstraction may be beyond the expertise of non-technical users --- and, further, those seeking to interact with such data may not be familiar with the nature of the systems involved. 

Again, standards will be important. 
Given that provenance information may be relevant for different audiences, there may be a need for various tools and extensions to assist. 
Some may describe lower-level technical details for a highly specialist audience, whereas others may, e.g., simply list the entities involved in order to assist a non-expert. 

There is a clear opportunity for human computer interaction (HCI) research to assist in ensuring that decision provenance approaches enable a representation that assists end-users regarding their accountability concerns. 
A number of techniques have been suggested as ways in which provenance data can be made more usable and understandable, including Natural Language Interfaces for Databases (NLIDBs)~\cite{li2012}, data visualisation techniques (such as graphs and plots)~\cite{chen2012, li2012, oliveira2017}, as well as using online games~\cite{bachour2015} and comics~\cite{schreiber2017} as a means for describing captured provenance information to end users. 
{
Generally, more work is required to explore the presentation of such data for accountability purposes, in ways that support the various perspectives (including users, technical experts, regulators, auditors, etc.) relevant in an accountability context. }

\section{Concluding remarks}

There are strong pressures for improving the levels of accountability for technology -- driven by societal demands, increasingly stringent legal requirements, and for reasons of public acceptance and adoption. In line with this, there is much discussion of algorithmic accountability, with a particular focus on automated decision-making (including ML). However, less-considered are the accountability challenges relating to the broader systems context -- particularly as systems are increasingly interconnected. Since these data-driven assemblages tend to be opaque, there is a pressing need to expose the nature and flow of data leading to, and resulting from, a decision or action to help raise levels accountability in interconnected environments.

We therefore propose decision provenance as an important piece of this accountability `jigsaw', offering much potential for assisting with issues of responsibility, technical compliance, and improved user agency in systems. 
Decision provenance won't itself solve all the complexities of increasingly interconnected systems, rather it provides visibility over the relations and dependencies of such systems where such visibility may not otherwise exist. As we have outlined, this has direct benefits from an accountability perspective.

In all, our wider goal is to highlight and raise awareness that exposing information of the contexts and nature  of data flow, and the broader decision pipeline, is important, relevant and complements other accountability work that focuses on the decision-making elements (e.g. ML models). 
Indeed, means for increasing accountability in a systems-of-systems context will only  increase in importance as visions of pervasive computing: smart cities, the Internet of Things, autonomous transport systems, etc., become a reality.
By presenting the concept of decision provenance, we seek to both focus and drive research efforts in the area, to realise the opportunities for provenance to assist systems accountability concerns.

\section*{Acknowledgment}

We acknowledge the financial support of the UK Engineering and Physical Sciences Research Council (EPSRC) [EP/P024394/1, EP/R033501/1], and also Microsoft, through the Microsoft Cloud Computing Research Centre.

\bibliographystyle{IEEEtran}

\newpage
\begin{IEEEbiography} [{\includegraphics[width=1in,height=1.25in,clip,keepaspectratio]{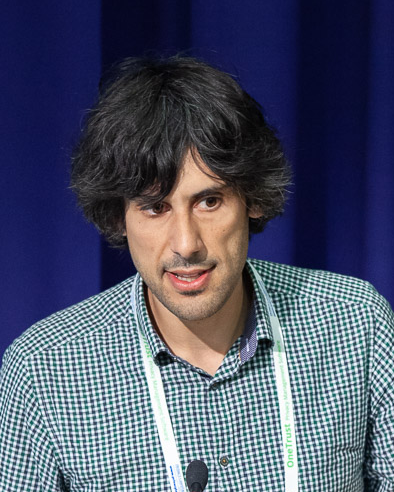}}]
{Jatinder Singh} is based at the Department of Computer Science and Technology (Computer Laboratory), University of Cambridge, where he leads the Compliant and Accountable Systems research group. The group focuses on the intersections of CS and law, looking at means and mechanisms for better aligning technology with legal concerns, and \textit{vice-versa}. From a technical perspective this work spans areas including security, privacy, data management, and auditing, typically in the context of cloud and distributed systems. He also co-chairs the Cambridge Trust \& Technology Initiative, and is a Fellow of the Alan Turing Institute, the UK's national centre for data science and artificial intelligence. He received his PhD in computer science from the University of Cambridge and has some background in law.
\end{IEEEbiography}

\begin{IEEEbiography}  [{\includegraphics[width=1in,height=1.25in,clip,keepaspectratio]{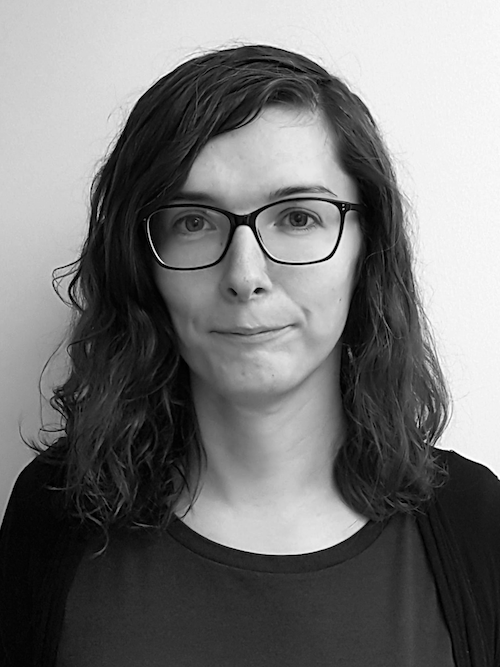}}]
{Jennifer Cobbe} is the Coordinator of the University of Cambridge's Trust \& Technology Initiative and a researcher in the Compliant and Accountable Systems research group in the Department of Computer Science \& Technology at Cambridge. She holds a PhD in Law and an LLM in Law and Governance from Queen's University, Belfast; for her PhD, she studied machine learning in commercial and state internet surveillance, data protection, and privacy.
\end{IEEEbiography}

\begin{IEEEbiography} [{\includegraphics[width=1in,height=1.25in,clip,keepaspectratio]{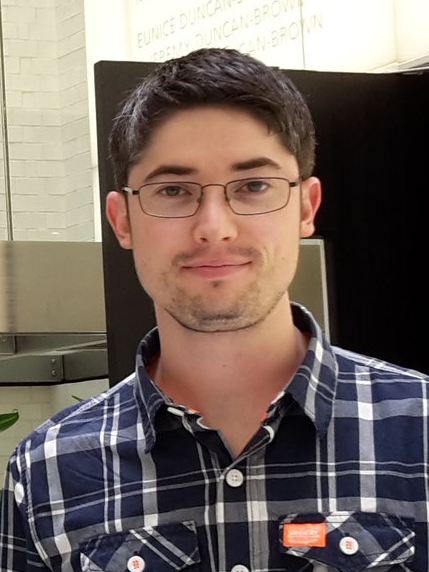}}]
{Chris Norval} is a postdoctoral researcher in the Compliant and Accountable Systems research group in the Department of Computer Science \& Technology at the University of Cambridge. He completed his PhD in Human Computer Interaction at the University of Dundee in 2014. He has since worked as a data analyst in the games industry, as well as investigating ethical issues of machine learning as a postdoctoral researcher at the University of St Andrews.
\end{IEEEbiography}

\EOD
\end{document}